\documentclass[twocolumn,showpacs,showkeys,superscriptaddress]{revtex4}
\usepackage{amsmath}
\usepackage{amsfonts}
\usepackage{amssymb}
\usepackage{color}

\begin{document}

{\raggedleft {\it Accepted for publication in Physical Review Letters}}
\\

\title{Generalized Magnetofluid Connections in Relativistic Magnetohydrodynamics}
\author{Felipe A. Asenjo}
\email{felipe.asenjo@uai.cl}
\affiliation{Facultad de Ingenier\'{\i}a y Ciencias, Universidad Adolfo Ib\'a\~nez, Santiago, Chile.}
\author{Luca Comisso}
\email{luca.comisso@polito.it}
\affiliation{Dipartimento Energia, Politecnico di Torino, Corso Duca degli Abruzzi 24, 10129, Torino, Italy.}
\affiliation{Istituto dei Sistemi Complessi - CNR, Via dei Taurini 19, 00185, Roma, Italy.}

%\date{\today}

\begin{abstract}
The concept of magnetic connections is extended to non-ideal relativistic magnetohydrodynamical plasmas. 
Adopting a general set of equations for relativistic magnetohydrodynamics including thermal-inertial, thermal electromotive, Hall and current-inertia effects, we derive a new covariant connection equation  showing the existence of generalized magnetofluid connections that are preserved during the dissipationless plasma dynamics.
These connections are intimately linked to a general antisymmetric tensor that unifies the electromagnetic and fluid fields, allowing the extension of the magnetic connection notion to a much broader concept.
\end{abstract}


\maketitle

In 1958 Newcomb \cite{Newcomb} showed that in a plasma that satisfies the ideal Ohm's law, two plasma elements connected by a magnetic field line at a given time will remain connected by a field line for all subsequent times. This occurs because the plasma moves with a transport velocity which preserves the ``magnetic connections" between plasma elements. This is one of the most fundamental and relevant ideas in plasma physics.

The conservation of the magnetic connections imposes significant constraints on the plasma dynamics. Moreover, it provides the very basis of central concepts in classical (neither relativistic nor quantum) plasma physics, such as magnetic field line motion, magnetic topology and magnetic reconnection. However, as the plasma energy increases (both in laboratory \cite{Nilson_2006,Li_2007,Ping_2014} and astrophysical contexts \cite{Tavani_2011,Abdo_2011,Zamaninasab_2014}), a generalization of these notions to the relativistic regime is required. Indeed, in a relativisic plasma different physical concepts must be taken into account,  such as the distinction between magnetic and electric fields, which are reference-frame dependent.

The key ideas to advance in the understanding of the magnetic connection notion in the relativistic regime have been first discussed  by Newcomb \cite{Newcomb}, and then thoughtfully investigated by Pegoraro  \cite{pegoraroEPJ}. In this latter work, using a covariant formulation for the dynamical equations of an ideal relativistic magnetohydrodynamic (MHD) plasma, it is shown that the magnetic connections are preserved by taking advantage of the antisymmetry of the electromagnetic field tensor $F^{\mu\nu}$. In this case,  the fundamental equation underlying the connection concept takes the form \cite{pegoraroEPJ}
\begin{equation}\label{pegoraroresult}
\frac{d}{d\tau}\left( dl_\mu F^{\mu\nu}\right)=-\left(\partial^\nu U_\beta\right)\left(dl_\alpha F^{\alpha\beta}\right)\, ,
\end{equation}
where $d/d\tau = U_\nu \partial^\nu$ is the convective derivative along the moving plasma  with four-velocity $U_\nu$ and satisfying the ideal Ohm's law $ U_\nu F^{\mu\nu}=0$. Here, $dl_\mu$ is the  four-vector that connects two plasma elements. From Eq. \eqref{pegoraroresult} it follows that if  $dl_\mu F^{\mu\nu}=0$ initially,
it will remain null for all subsequent times. This generalizes the concept of magnetic connections, which can be recovered if one consider simultaneous events \cite{pegoraroEPJ}.

Despite these substantial advances in the understanding of the relativistic plasma dynamics, one should recall that there are plenty of non-ideal effects which play crucial roles in the dynamics of laboratory and astrophysical plasmas. In a non-ideal relativistic MHD plasma, effects such as thermal-inertial, Hall, or current inertia effects, modify the equation of motion and Ohm's law. Therefore Eq. \eqref{pegoraroresult} is no longer valid and so the magnetic connections are not conserved \cite{Nilson_2006,Li_2007,Ping_2014,Zenitani2011,ComissoAsenjo2014}. However, one might expect that in these dissipationless cases the magnetic connections could be replaced by more sofisticated ``generalized connections''.  Along this work, we show that these generalized connections indeed exist.

We suggest that the nature of the new connections can be understood through the emergence of a general antisymmetric tensor field, which can be constructed as a combination of several physical quantities of the MHD plasma.
This insight is deeply inspired by the ideas proposed by Bekenstein \cite{Bekenstein_1987} and Mahajan \cite{mahajanU}, who have shown that the helicity invariants characterizing the ideal plasma dynamics can be generalized for relativistic plasmas by realizing that fluid and electromagnetic fields couple in one antisymmetric tensor. Similar ideas have been also successfully applied to develop relativistic theories of vorticity generation \cite{mahajan2010,asenjoGR,Gao_2014},  to study the topological constraints imposed by the plasma helicity \cite{Yoshida_2014,Yoshida2_2014}, and in the construction of generalized vorticity invariants in the framework of non-abelian plasmas \cite{Bambah_2006} and spinning quantum plasmas \cite{MahajanAsenjoPRL2011}.

In our analysis we intend to extend the connection concept by considering physical contents beyond the ideal MHD description. 
To this aim, we consider a plasma governed by the generalized relativistic MHD equations recently derived by Koide~\cite{koide}. These equations retain many effects generally neglected in previous simpler models \cite{lich,Anile_1989}, such as thermal-inertial effects, thermal electromotive effects, current inertia effects and Hall effects. 
The spacetime is assumed to be flat and defined by the Minkowski metric tensor $\eta_{\mu\nu} = {\rm{diag}}(-1,1,1,1)$. 
For an electron-ion plasma with density $n$, charge density $q=ne$, normalized four-velocity $U^\mu$ (such that $U_\mu U^\mu=-1$) and normalized four-current density $J^\mu$, the equations that govern the dynamics of the plasma are the continuity equation $\partial_\mu (qU^\mu)=0$, the generalized momentum equation
\begin{equation}\label{momentumEquation}
  \partial_\nu\left(hU^\mu U^\nu+\frac{\mu h}{q^2}J^\mu J^\nu\right)=-\partial^\mu p+J_\nu F^{\mu\nu}\, ,
\end{equation}
and the generalized Ohm's law
\begin{eqnarray}\label{ohmlaw1orig}
 &&\partial_\nu\left[\frac{\mu h}{q}(U^\mu J^\nu+J^\mu U^\nu)-\frac{\mu\Delta\mu h}{q^2}J^\mu J^\nu \right]\nonumber\\
 &&\qquad=\frac{1}{2}\partial^\mu \Pi+qU_\nu F^{\mu\nu}-\Delta\mu J_\nu F^{\mu\nu}+q R^\mu\, .
\end{eqnarray}
Here, $h$ denotes the MHD enthalpy density, while $\Pi=p\Delta\mu -\Delta p$, with 
 $p=p_+ + p_-$ and $\Delta p = p_+ - p_-$ denoting the total pressure and the pressure difference between the fluids, respectively. 
We indicate with the subscript $+$ ($-$) the positively (negatively) charged fluid. 
The reduced (rest) mass can be recognized as $\mu=m_+ m_-/m^2$, with $m=m_+ + m_-$, wherease $\Delta\mu=(m_+ - m_-)/m$. 
The frictional four-force density between the fluids is
\begin{equation}\label{}
  R^\mu=-\eta\left[J^\mu+Q(1+\Theta)U^\mu\right]\, ,
\end{equation}
where $\Theta$ is the thermal energy exchange rate from the negatively to the positively charged fluid, $\eta$ is the plasma resistivity, and $Q=U_\mu J^\mu$.
In Eq. \eqref{momentumEquation}, the current inertia effects arise from
the left-hand side. Similarly,  the thermal electromotive effects appear as inertial effects corrections in the left-hand side of Eq. \eqref{ohmlaw1orig}, whereas in the right-hand side it is taken into account the contributions of the  thermal electromotive force and the Hall effect.

As usual, $F^{\mu \nu} = {\partial^\mu}{A^\nu} - {\partial^\nu}{A^\mu}$ is the electromagnetic field tensor ($A^\mu$ is the four-vector potential), which obeys Maxwell's equations
\begin{equation}\label{}
  \partial_\nu F^{\mu\nu}=4\pi J^\mu\, ,\qquad
 \partial_\nu F^{*\mu\nu}=0\, .
\end{equation}
Of course, $F^{*\mu\nu} = (1/2) \epsilon^{\mu\nu\alpha\beta} F_{\alpha\beta}$ is the dual of $F^{\mu\nu}$, and $\epsilon^{\mu\nu\alpha\beta}$ indicates the Levi-Civita symbol.

Although the inclusion of different relativistic effects complicates the MHD equations and a generalization of Newcomb's results seems difficult to achieve, a hint may be gleaned directly from Pegoraro's analysis \cite{pegoraroEPJ}, where the antisymmetric form of tensors is what allows to prove the existence of preserved magnetic connections. 
Therefore, the initial step of our search for generalized connections consists in re-formulating the generalized Ohm's law \eqref{ohmlaw1orig} in a form suitable to be antisymmetrized. We will see (below) that to achieve this, it is imperative to introduce an antisymmetric tensor that unifies electromagnetic and fluid fields.

First, using the continuity equation $\partial_\mu (qU^\mu)=0$, and the charge conservation equation $\partial_\mu J^\mu=0$ (coming from Maxwell's equations), we can rewrite the generalized Ohm's law \eqref{ohmlaw1orig} as
\begin{eqnarray}\label{apeOhm1}
  &&J_\nu\partial^\nu\left(\frac{\mu h}{q}U^\mu\right)+qU_\nu\partial^\nu\left(\frac{\mu h}{q^2} J^\mu \right)-\Delta\mu J_\nu\partial^\nu\left(\frac{\mu h}{q^2}J^\mu\right)\nonumber\\
  &&\qquad\quad=\frac{1}{2}\partial^\mu \Pi+qU_\nu F^{\mu\nu}-\Delta\mu J_\nu F^{\mu\nu}+q R^\mu\, .
\end{eqnarray}
Then,  in analogy with the definition of the electromagnetic field tensor, we construct an antisymmetric flow-field tensor
\begin{eqnarray}\label{fluidVortic}
  S^{\mu\nu}& = &\partial^\mu\left(\frac{h}{q}U^\nu\right)-\partial^\nu\left(\frac{h}{q}U^\mu\right) \, ,
\end{eqnarray}
and an antisymmetric current-field tensor
\begin{eqnarray}\label{}
  \Lambda^{\mu\nu}& = &\partial^\mu\left(\frac{h}{q^2}J^\nu\right)-\partial^\nu\left(\frac{h}{q^2}J^\mu\right) \, .
\end{eqnarray}
The antisymmetric tensor \eqref{fluidVortic} represents the covariant generalization of the fluid vorticity
and it has been previously introduced for relativistic hot plasmas in flat spacetimes \cite{mahajanU,mahajan2010} and in general relativity \cite{asenjoGR}.
In both  above tensors, the four-vectors $h U^\nu /q$ and $h J^\nu /q^2$ may be viewed as the potentials (equivalent to the four-vector potential $A^\mu$) for the flow-field and the current-field, respectively. With these definitions Eq. \eqref{apeOhm1} becomes
\begin{eqnarray}\label{}
&&\mu J_\nu\partial^\mu\left(\frac{ h}{q}U^\nu\right)+\mu qU_\nu\partial^\mu\left(\frac{h}{q^2} J^\nu \right)=\nonumber\\
&&\quad\frac{1}{2}\partial^\mu \Pi+qU_\nu \left(F^{\mu\nu}+\mu \Lambda^{\mu\nu}\right)\nonumber\\
&&\quad- J_\nu \left( \Delta\mu F^{\mu\nu}-\mu S^{\mu\nu}+\mu\Delta\mu \Lambda^{\mu\nu}\right)+q R^\mu\, .
\end{eqnarray}
To move forward in the antisymmetrization scheme, we define the general transport four-velocity
\begin{equation}\label{TransVelo}
  {\cal U}^\mu = U^\mu-\frac{\Delta\mu}{q}J^\mu\, ,
\end{equation}
which allow us to put the generalized Ohm's law in the form
\begin{eqnarray}\label{}
&&{\mu Q}\left[\frac{1}{q}\partial^\mu\left(\frac{h}{q}\right)+\partial^\mu\left(\frac{h}{q^2}\right)\right]+\frac{\mu h}{q^2}\partial^\mu Q=\nonumber\\
&&\quad\frac{1}{2q}\partial^\mu \Pi+{\cal U}_\nu \left(F^{\mu\nu}+\mu \Lambda^{\mu\nu}\right)+\frac{\mu}{q}J_\nu  S^{\mu\nu}+ R^\mu\, .
\end{eqnarray}
Finally, we can manipulate the left-hand side of this equation to reformulate the generalized Ohm's law in the concise and elegant form
\begin{equation}\label{Eq5Ap}
 {\Sigma}^\mu={\cal U}_\nu {\cal M}^{\mu\nu}+ {R}^\mu\, ,
\end{equation}
where we have defined the antisymmetric tensor
\begin{eqnarray}\label{magnetofluidUnifiedTensor}
  {{\cal M}^{\mu \nu}}={F}^{\mu\nu}-\frac{\mu}{\Delta\mu}{W}^{\mu\nu}\, ,
\end{eqnarray}
with the antisymmetric general flow-field tensor
\begin{equation}\label{}
{W}^{\mu\nu}={S}^{\mu\nu}-\Delta\mu {\Lambda}^{\mu\nu}=\partial^\mu\left(\frac{h}{q}{\cal U}^\nu\right)-\partial^\nu\left(\frac{h}{q}{\cal U}^\mu\right)\, ,
\end{equation}
and $\Sigma^\mu= \partial^\mu\left[{\mu h Q}/{q^2}+{\mu h}/({q\Delta\mu})\right]+({\mu}/{\Delta\mu})\chi^\mu$, with
\begin{equation}
\chi^\mu=U_\nu\partial^\nu\left(\frac{h}{q}U^\mu\right)+\frac{\Delta\mu Q}{q}\partial^\mu\left(\frac{h}{q}\right)-\frac{\Delta\mu}{2\mu q}\partial^\mu\Pi\, .
\end{equation}

The tensor ${{\cal M}^{\mu \nu}}$ defined in Eq.~\eqref{magnetofluidUnifiedTensor} is a {\it generalized magnetofluid field tensor}. This quantity
represents an effective field tensor which unifies the electromagnetic and fluid forces (through $F^{\mu\nu}$ and $W^{\mu\nu}$). 
Note that the tensor $W^{\mu\nu}$ contains the information of  both the flow and current fields.
We will show later that the effective field tensor ${\cal M}^{\mu\nu}$ is crucial in revealing fundamental properties of the system. This tensor is a more complex version  of the unified magnetofluid field tensor introduced in Ref.~\cite{mahajanU} for the simpler case of one-species hot relativistic plasmas.

To find the evolution of the generalized magnetofluid field tensor, we have to take  the curl of Eq.~\eqref{Eq5Ap}. Applying first $\epsilon_{\alpha\beta\gamma\mu}\partial^\gamma$ and then $\epsilon^{\alpha\beta\lambda\phi}$ to Eq.~\eqref{Eq5Ap}, we obtain the equation
\begin{equation}\label{Important}
  \frac{d {\cal M}^{\lambda\phi}}{d\tau}=\partial^\lambda {\cal U}_\nu {\cal M}^{\phi\nu}-\partial^\phi {\cal U}_\nu {\cal M}^{\lambda\nu}-\frac{\mu}{\Delta\mu}{\cal Z}^{\lambda\phi} +\partial^\lambda {R}^\phi-\partial^\phi {R}^\lambda \, ,
\end{equation}
where now we have defined a general convective derivative $d/d\tau = {\cal U}_\nu\partial^\nu$, and introduced the antisymmetric tensor
\begin{equation}\label{eqZ}
  {\cal Z}^{\lambda\phi}=\partial^\lambda\chi^\phi- \partial^\phi\chi^\lambda\, .
\end{equation}
Therefore, Eq.~\eqref{Important} is an antisymmetric form of the generalized Ohm's law.

Before proceeding further in showing the existence of generalized magnetofluid connections, let us analyze the physical content of the tensor field ${\cal Z}^{\lambda\phi}$. To this aim, we rewrite Eq.~\eqref{momentumEquation} with the help of the continuity equation as
\begin{equation}\label{momentumEquation2}
  U_\nu \partial^\nu\left(\frac{h}{q}U^\phi\right)=-\frac{\mu}{q} J^\nu \partial_\nu\left(\frac{h}{q^2} J^\phi\right)-\frac{1}{q}\partial^\phi p+\frac{1}{q}J_\nu F^{\phi\nu} \, ,
\end{equation}
and then we use it to cast Eq.~\eqref{eqZ} in a sum of different relativistic contributions to Ohm's law
\begin{equation}\label{}
 {\cal Z}^{\lambda\phi}={\cal Z}^{\lambda\phi}_h+{\cal Z}^{\lambda\phi}_p+{\cal Z}^{\lambda\phi}_H+{\cal Z}^{\lambda\phi}_c \, ,
\end{equation}
where
\begin{eqnarray}\label{contributionsZ}
  {\cal Z}_h^{\lambda\phi}&=&\Delta\mu\left[\partial^\lambda\left(\frac{Q}{q}\right)\partial^\phi\left(\frac{h}{q}\right)-\partial^\phi\left(\frac{Q}{q}\right)\partial^\lambda\left(\frac{h}{q}\right)\right]\, ,\nonumber\\
  {\cal Z}_p^{\lambda\phi}&=&\frac{\partial^\lambda q}{q^2}\partial^\phi\left( p+\frac{\Delta\mu}{2\mu} \Pi\right)-\frac{\partial^\phi q}{q^2}\partial^\lambda\left( p+\frac{\Delta\mu}{2\mu}\Pi\right)\, ,\nonumber\\
  {\cal Z}_H^{\lambda\phi}&=&\partial^\lambda\left(\frac{1}{q}J_\nu F^{\phi\nu}\right)-\partial^\phi\left(\frac{1}{q}J_\nu F^{\lambda\nu}\right)\, ,\nonumber\\
  {\cal Z}_c^{\lambda\phi}&=&  -\partial^\lambda \left[\frac{\mu}{q} J^\alpha\partial_\alpha\left(\frac{h}{q^2}J^\phi\right)  \right]+\partial^\phi \left[\frac{\mu}{q} J^\alpha\partial_\alpha\left(\frac{h}{q^2}J^\lambda\right)  \right]\, .\nonumber\\
&&
\end{eqnarray}
The antisymmetric tensors ${\cal Z}_h^{\lambda\phi}$ and ${\cal Z}_p^{\lambda\phi}$ are due to the thermal-inertial and thermal electromotive effects of the MHD plasma. In particular, ${\cal Z}_p^{\lambda\phi}$ contains the pressure contribution, which is absent if the equation of state of the MHD plasma is such that $p, \Delta p\propto q$. The contributions coming from the Hall effect in the generalized Ohm's law are instead retained by the tensor ${\cal Z}_H^{\lambda\phi}$, while ${\cal Z}_c^{\lambda\phi}$ appears owing to current inertia effects.

We are now able to demonstrate that  when the frictional four-force density $R^\mu$ is negligible, e.g. when the evolution of the system is fast compared to the dissipation time scale, there exist generalized magnetofluid connections that are preserved during the dynamics of the relativistic plasma.

With this aim in mind, we define a general displacement four-vector $\Delta x^\mu$ of a general element that is transported by the  general four-velocity
\begin{equation}\label{dX}
  \frac{\Delta x^\mu}{\Delta\tau} = {\cal U}^\mu+\frac{\mu}{\Delta\mu} {\cal D}^\mu\, ,
\end{equation}
where $\Delta\tau$ is the variation of the proper time and ${\cal D}^\mu$ is a four-vector field which satisfies the equation 
\begin{equation}\label{eqD}
  {\cal M}^{\nu\phi}\partial^\lambda {\cal D}_\nu-  {\cal M}^{\nu\lambda}\partial^\phi {\cal D}_\nu={\cal Z}^{\lambda\phi}\, .
\end{equation}
The four-vector ${\cal D}^\mu$ contains all the (inertial-thermal-current-Hall) information of ${\cal Z}^{\mu\nu}$, and it turns out to be essential to prove the existence of the generalized connections.
Similarly to Ref.~\cite{pegoraroEPJ}, we introduce the space-like event-separation four-vector $d l^\mu=x'^{\mu}-x^{\mu}$ between two different elements. Two events are simultaneous in a frame where $dl^0=0$.
From the definition of $dl^\mu$ and the general four-velocity defined in Eq.~\eqref{dX}, it follows that this four-vector is transported according to $(d/d\tau)dl^\mu={\cal U}'^\mu+({\mu}/{\Delta\mu}){\cal D}'^\mu-{\cal U}^\mu-({\mu}/{\Delta\mu}){\cal D}^\mu={\cal U}^\mu(x_\alpha+dl_\alpha)+({\mu}/{\Delta\mu}){\cal D}^\mu(x_\alpha+dl_\alpha)-{\cal U}^\mu(x_\alpha)-({\mu}/{\Delta\mu}){\cal D}^\mu(x_\alpha)$. Therefore, the four-vector $dl^\mu$ fulfills 
\begin{equation}\label{dldt}
  \frac{d }{d\tau}dl^\mu=dl^\alpha\partial_\alpha\left({\cal U}^\mu+\frac{\mu}{\Delta\mu}{\cal D}^\mu \right)\, .
\end{equation}
With the help of Eq.~\eqref{dldt}, and neglecting the frictional four-force density $R^\mu$ in Eq.~\eqref{Important}, we can calculate
\begin{eqnarray}
  \frac{d }{d\tau}\left(dl_\lambda {\cal M}^{\lambda\phi}\right)
   &=&-\partial^\phi {\cal U}_\nu \left(dl_\lambda {{\cal M}}^{\lambda\nu}\right)\nonumber\\
   &&+\frac{\mu}{\Delta\mu} dl_\lambda\left(\partial^\lambda {\cal D}_\nu {\cal M}^{\nu\phi}-{\cal Z}^{\lambda\phi}\right)\, .
\end{eqnarray}
This equation, which is in terms of the new four-vector field ${\cal D}^\mu$, can finally be rewritten using Eq.~\eqref{eqD} as
\begin{equation}\label{CTequation}
  \frac{d }{d\tau}\left(dl_\lambda {\cal M}^{\lambda\phi}\right) =-\left(dl_\lambda {{\cal M}}^{\lambda\nu}\right)\partial^\phi\left({\cal U}_\nu+\frac{\mu}{\Delta\mu}  {\cal D}_\nu \right)\, .
\end{equation} 
This is a crucial equation which reveals the existence of generalized magnetofluid connections that are preserved during the plasma dynamics. Indeed, from this equation it follows that if $dl_\lambda {\cal M}^{\lambda\phi}=0$ initially, then $d/d\tau (dl_\lambda {\cal M}^{\lambda\phi})=0$ for every time, and so $dl_\lambda {\cal M}^{\lambda\phi}$ will remain null at all times. Of course, regularity properties of the velocity field \eqref{dX} are assumed.

The ``magnetofluid connection equation'' \eqref{CTequation} generalizes Eq.~\eqref{pegoraroresult} for a relativistic electron-ion MHD plasma with thermal-inertial, Hall, thermal electromotive and current inertia effects. In particular, since 
\begin{equation}\label{generalizedmagneticconnectionsM}
d{l_\lambda }{{\cal M}^{\lambda \phi }} = d{l_\lambda }{F^{\lambda \phi }} - \frac{\mu }{{\Delta \mu }}d{l_\lambda }{{W}^{\lambda \phi }} \, ,
\end{equation}
the generalized magnetofluid connections reduce to the well-known magnetic connections if $(\mu/\Delta\mu)d{l_\lambda }{{W}^{\lambda \phi }} \to 0$ and $dl_0=0$. 
On the contrary, in the case $F^{\lambda \phi} \to 0$, $W^{\lambda \phi } \to S^{\lambda \phi}$ and $dl_0=0$, the preserved connections are those of the fluid vorticity.

Note that the four-vector connections \eqref{generalizedmagneticconnectionsM} are transported by the general four-velocity ${\cal U}_\mu+(\mu/\Delta\mu) {\cal D}_\mu$. The four-vector ${\cal D}^\mu$ is fundamental to prove the existence of the generalized connections and it can be found by solving the differential equation
\begin{equation}\label{solutD}
\partial^\mu {\cal D}_\nu= {\cal N}_{\phi\nu}\partial^{\mu}\chi^\phi\, ,
\end{equation}
obtained from Eq.~\eqref{eqD}. Here, ${\cal N}_{\mu\nu}$ is the inverse of the ${\cal M}_{\mu\nu}$ matrix (${\cal M}^{\mu\alpha}{\cal N}_{\alpha\nu}={\delta^\mu}_\nu$). The field ${\cal N}_{\mu\nu}$ exists since ${\cal M}_{\mu\nu}$ is non-sigular for non-ideal MHD. To show this, it is convenient to separate Eq.~\eqref{Eq5Ap} into time and spatial components. Defining the components of a generalized electric-like field ${\cal E}^i={\cal M}^{0i}$ and a generalized magnetic-like field ${\cal B}^k=(1/2)\epsilon^{ijk}{\cal M}_{ij}$, the determinant of ${\cal M}_{\mu\nu}$ is $\left\| {\cal M} \right\| =\left(\vec{\cal E}\cdot\vec{\cal B}\right)^2$. On the other hand, the time component of  Eq.~\eqref{Eq5Ap} becomes $\vec {\cal U}\cdot\vec{\cal E}=\Sigma^0-R^0$, whereas the spatial components produce the equation ${\cal U}^0\vec{\cal E}+\vec{\cal U}\times\vec{\cal B}=\vec{\Sigma}-\vec{R}$, implying that ${\cal U}^0 \vec{\cal E}\cdot \vec{\cal B}=\left(\vec\Sigma-\vec R\right)\cdot\vec{\cal B}$.
Therefore, as long as $\vec{\cal E}\cdot\vec{\cal B} = \vec{\Sigma}\cdot\vec{\cal B}/{\cal U}^0 - \vec{R}\cdot\vec{\cal B}/{\cal U}^0 \neq 0$, the tensor ${\cal N}_{\mu\nu}$ exists (i.e., $\left\| {\cal M} \right\| \neq 0$). In the particular case of vanishing resistivity, it can be $\vec{\cal E}\cdot\vec{\cal B} \neq 0$ due to the non-ideal effects in $\vec{\Sigma}$.

We also observe that in general, if there is simultaneity bewteen two events, the condition  $d{l_\lambda }{{\cal M}^{\lambda \phi }} = 0$ can be written as the vectorial conditions 
\begin{equation}\label{}
d\vec l\cdot \vec {\cal E}=0  \quad {\text{and}}\quad  d\vec l\times \vec {\cal B}=0 \, .
\end{equation}
On the other hand, when the events are not simultaneous, Pegoraro has shown that simultaneity can be recovered resetting the time \cite{pegoraroEPJ}. This can be achieved projecting the new trajectories of the fluid elements  in a 3D space by changing $dl_\mu\rightarrow d{l'}_\mu=dl_\mu+({\cal U}_\mu + \mu H_\mu/\Delta\mu) d\lambda$,
such that in this new reference frame $dl'_0=0$. This allows to have unaltered generalized connections whenever $H_\mu$ fulfills the generalized Ohm's law \eqref{Eq5Ap}.  Thus, a possible solution can be expressed as $H_\alpha=(\Delta\mu/\mu){\cal N}_{\mu\alpha}\Sigma^\mu={\cal N}_{\mu\alpha}\chi^\mu+{\cal N}_{\mu\alpha}\partial^\mu\left({\Delta\mu h Q}/{q^2}+{h}/{q}\right)$. Notice the resemble with the ${\cal D}_\alpha$ field, which can be calculated from Eq.~\eqref{solutD} as 
\begin{equation}
{\cal D}_\alpha={\cal N}_{\mu\alpha}\chi^\mu+\varepsilon_\alpha\, ,
\end{equation}
with $\partial^\mu\varepsilon_\nu= \partial^\mu {\cal N}_{\nu\beta}\chi^\beta$.

Our procedure has lead us to a completely general result, and of course, also nonrelativistic systems are covered.  As an illustration, and for the sake of simplicity, let's consider a non-relativistic Hall MHD plasma. For this case, we neglect the thermal-inertial and thermal electromotive effects. In the non-relativistic limit $J_\mu  {W}^{\mu\nu}={\mathcal O}(v^2)$ and ${\cal Z}^{\lambda\phi}_c$ is negligible. Thus, the only contribution of the ${\cal Z}^{\lambda\phi}$ field is ${\cal Z}^{\lambda\phi}_H$, implying that $\chi^\phi=J_\mu F^{\phi\mu}/q\approx J_\mu {\cal M}^{\phi\mu}/q$  because of our approximations. Therefore, Eq.~\eqref{solutD} becomes
$\partial^\mu {\cal D}_\nu\approx -\partial^\mu\left(J_\nu/q\right)+{\cal N}_{\phi\nu}\partial^\mu F^{\phi\alpha}J_\alpha/q$.
A simple analytical solution of this equation may be obtained for quasi-uniform electromagnetic fields. Under this approximation, we have ${\cal D}_\nu\approx - J_\nu /q$. Substituting this solution into Eq.~\eqref{dX}, we can see that in this limit, the general vectorial velocity
\begin{equation}\label{elecVel2}
\vec{\cal U}+\frac{\mu}{\Delta\mu}\vec{\cal D} \approx \vec v -\frac{1}{q}\left(\Delta\mu+\frac{\mu}{\Delta\mu}\right)\vec J\, 
\end{equation}
preserves the topology of the non-relativistic limit (with no thermal-inertial effects) of the generalized magnetic-like field
\begin{equation}\label{topoB2}
\vec{\cal B} \approx \vec B-\frac{\mu}{\Delta\mu} \nabla \times \left(\vec v-\frac{\Delta\mu}{q}\vec J\right)\, .
\end{equation} 
Therfore, in the considered limit the preserved connections are related to the rotation of the fluid electron canonical momentum, as shown by Pegoraro \cite{pegoraronpg}. 
We also note that it is common to assume $m_+\gg m_-$, which yields $\mu\approx m_-/m_+\ll 1$ and $\Delta\mu\approx 1$. In this case the transport velocity \eqref{elecVel2} becomes the electron velocity, while the preserved connections are simply the magnetic connections.

A difference of the theory developed here with the ideal MHD connection theorem  \cite{Newcomb,pegoraroEPJ} is worth to be mentioned. In the ideal case, the magnetic connection concept is well defined when $|\vec{E}| < |\vec{B}|$ (where $U_\nu$ is time-like). On the contrary, when $|\vec{E}| > |\vec{B}|$, which may occur in extreme astrophysical environments \cite{amano, koma}, the MHD approximation breaks down since $U_\nu$ becomes space-like \cite{Newcomb}. In the non-ideal case treated here, the same analysis is not directly applicable. The field strengths to be compared in the non-ideal system are $|\vec{\cal E}|$ and $|\vec{\cal B}|$ instead of $|\vec{E}|$ and $|\vec{B}|$. In this case, the generalized Ohm's law \eqref{Eq5Ap} gives that $\vec{\cal U}={\cal U}^0 (\vec{\cal E}\times\vec{\cal B})/|\vec{\cal B}|^2+(\vec{\cal U}\cdot\vec{\cal B}/|\vec{\cal B}|^2)\vec{\cal B}+\vec{\cal B}\times\vec\Sigma$, when the frictional four-force density is neglected. Therefore, since the fields $\vec{\cal B}$ and $\vec{\cal E}$ depend on the  general transport four-velocity, it is not straightforward to establish when the general transport velocity $\vec{\cal U}$ is time-like or space-like. However, as long as ${\cal U}_\nu$ is time-like, the generalized magnetofluid connection concept is well defined. In this case it is also possible to have generalized magnetic-like field connections $\vec{\cal B}$ in a frame where the magnetic field is null ($\vec{B}=0$) because of the current and velocity fields.

The generalization of the connection concept and of the connection equation to non-ideal relativistic MHD plasmas provides a strong theoretical framework for investigating high-energy plasmas. 
In this extended framework, the generalized magnetofluid field tensor \eqref{magnetofluidUnifiedTensor} and the general four-velocity \eqref{dX} play the role that the electromagnetic field tensor and the fluid four-velocity have in the ideal relativistic MHD description. Thereby, when dissipationless non-ideal effects are included, the preserved connections are no longer related to the electromagnetic field tensor ${F^{\mu \nu}}$ but to the magnetofluid field tensor ${{\cal M}^{\mu \nu}}$. This implies that it is possible to have reconnection of the magnetic field lines $\vec B$ while the generalized magnetic-like field lines $\vec{\cal B}$ remain conserved (in a frame where $dl_0=0$).

The conservation of these generalized magnetofluid connections profoundly affect the relativistic plasma dynamics by forbidding transitions between configurations with different magnetofluid connectivity. Also, because of these constraints, they could be crucial to understand the formation of small scale structures resulting from a complex nonlinear dynamics. Therefore, it would not be surprising that the ideas presented in this letter will allow us to gain a more detailed comprehension of relativistic phenomena in  plasmas. 
Furthermore, from the computational point of view, these ideas can be applied to verify the accuracy of relativistic MHD numerical simulations, using the magnetofluid connection equation \eqref{CTequation} as guidance for reliability.

Finally, the results presented here tell us that the connection idea, conceived first by Newcomb, could be a broader concept that permeates the entire energy scale in plasma physics.

{\it Acknowledgments.} F. A. A. thanks Fondecyt-Chile for Funding No. 11140025. L. C. thanks Xinjuan Tian for all her support.

\end{document}